\documentclass[prl,aps,twocolumn,showpacs]{revtex4}

\usepackage{amsgen,amstext,amsbsy,amsopn,amsthm, amssymb}
\usepackage{amsmath}

\newtheorem*{thm}{Theorem}

\theoremstyle{definition}

\newcommand{\beq}{\begin{equation}}
\newcommand{\eeq}{\end{equation}}
\newcommand{\beqa}{\begin{eqnarray}}
\newcommand{\eeqa}{\end{eqnarray}}

\newcommand{\half}{\mbox{$\frac{1}{2}$}}
\newcommand{\E}{\mathcal{E}}

\newcommand{\x}{{\bf x}}

\begin{document}


\title{One-dimensional Bosons in Three-dimensional Traps}

\author{Elliott H.~Lieb}
\email{lieb@princeton.edu}
\affiliation{Department of Physics,
Jadwin Hall, Princeton University,
P.~O.~Box 708, Princeton, New Jersey 08544}
\author{Robert Seiringer}
\altaffiliation{On leave from Institut f\"ur Theoretische Physik,
Universit\"at
Wien, Boltzmanngasse 5, A-1090 Vienna, Austria}
\email{rseiring@princeton.edu}
\affiliation{Department of Physics,
Jadwin Hall, Princeton University,
P.~O.~Box 708, Princeton, New Jersey 08544}
\author{Jakob Yngvason}
\email{yngvason@thor.thp.univie.ac.at}
\affiliation{Institut f\"ur Theoretische Physik,
Universit\"at
Wien, Boltzmanngasse 5, A-1090 Vienna, Austria}

\date{August 29, 2003}

\begin{abstract}
Recent experimental and theoretical work has indicated conditions in
which a trapped, low-density Bose gas ought to behave like the 1D
delta-function Bose gas solved by Lieb and Liniger.  Up to now the
theoretical arguments have been based on variational - perturbative
ideas or numerical investigations.  There are 4 parameters: density, transverse and longitudinal dimensions, and
scattering length.  In this paper we explicate 5 parameter regions in
which various types of 1D or 3D behavior occur in the ground state.  Our treatment is
based on a rigorous analysis of the many-body Schr\"odinger equation.

This version of the article is a combination of the earlier 4-page arXiv version and the revised 4-page Physical Review Letters version.  

\end{abstract}

\maketitle

It appears to be possible to do experiments in highly elongated traps
on ultra-cold Bose gases that are effectively 1D. More
precisely, the 1D Bose gas with delta function 2-body interaction, analyzed
long ago
\cite{LL,L}, should be visible, with its quasi-fermionic behavior
\cite{gir}, the absence of Bose-Einstein condensation (BEC) in a
dilute limit \cite{Lenard,pita,girardeau}, and an excitation
spectrum different from that predicted by Bogoliubov's theory 
\cite{L,jackson,komineas}.  Several theoretical investigations on the
transitions from 3D to an effective 1D behavior were triggered by
\cite{olshanii}. See, e.g., 
\cite{dunjko,petrov,kolomeisky,tanatar,girardeau2,das,das2, menotti}. 
Systems showing indications of such a transition have recently been
prepared experimentally \cite{bongs,goerlitz, greiner, schreck}.

The theoretical work on the dimensional cross-over for the ground
state in elongated traps has so far been based either on variational
calculations, starting from a 3D delta-potential
\cite{olshanii,das2,girardeau2}, or on numerical Quantum Monte Carlo
studies \cite{blume,astra} with more realistic, genuine 3D potentials,
but particle numbers limited to the order of 100.  This work is
important and has led to valuable insights, in particular about
different parameter regions \cite{petrov,dunjko}, but a more thorough
theoretical understanding is clearly desirable since this is not a
simple problem.  In fact, it is evident that for a potential with a
hard core the true 3D wave functions do not approximately factorize in
the longitudinal and transverse variables (otherwise the energy would
be infinite) and the effective 1D potential can not be obtained by
simply integrating out the transverse variables of the 3D
potential (that would immediately create an impenetrable barrier in
1D).  It is important to be able to demonstrate rigorously, and
therefore unambiguously, that the 1D behavior really follows from the
fundamental Schr\"odinger equation.  It is also important to
delineate, as we do here, precisely what can be seen in the different
parameter regions.  The full proofs of our assertions are long and
will be given elsewhere \cite{LSY}, but we emphasize that everything
can be rigorously derived from first principles.  In this paper we
state our main results and outline the basic ideas for the proofs.

We shall always be concerned with the ground state and with large particle number, 
$N \gg 1$, which is appropriate for the consideration of actual 
experiments.  In order to make precise statements we
take the limit $N\to\infty$ but the reader can confidently apply these 
limiting statement to finite numbers like $N=100$. 
Other parameters of the problem are the
scattering length, $a$, of the two-body interaction potential, $v$,
and two lengths, $r$ and $L$, describing the transverse and the
longitudinal extension of the trap potential, respectively.

It is convenient to write the
Hamiltonian in the following way (in units where $\hbar=2m =1$): 
\begin{align} \label{3dham}
H_{N,L,r,a}&=\sum_{j=1}^N \left( -\nabla^2_j +
V^{\perp}_{r}(\x^\perp_j)  + V_{L} (z_j) \right)\notag \\
 &+ \sum_{1\leq i<j\leq
N} v_{a}(|\x_i-\x_j|) 
\end{align}
with $\x=(x,y,z)=(\x^\perp,z)$ and with
\begin{align}
V^{\perp}_{r}(\x^\perp)&=\frac 1{r^2}
 V^{\perp}(\x^\perp/r)\ , \notag \\
V_L(z)=\frac 1{L^2} V (z/L)&\ , \quad v_{a}(|\x|)=\frac 1{a^2}v(|\x|/a)\ .
\end{align}
Here, $r, L, a$ are variable scaling parameters while $V^{\perp}$, $V$
and $v$ are fixed.  The interaction potential $v$ is supposed to be
nonnegative, of finite range and have scattering length 1; the scaled
potential $v_{a}$ then has scattering length $a$.  The external trap
potentials $V$ and $V^\perp$ confine the motion in the longitudinal
($z$) and the transversal ($\x^{\perp}$) directions, respectively, and
are assumed to be continuous and tend to $\infty$ as $|z|$ and $|
\x^{\perp}|$ tend to $\infty$.  To simplify the discussion we find it
also convenient to assume that $V$ is homogeneous of some order $s>0$, namely
$V(z)=|z|^s$, but weaker assumptions, e.g. asymptotic homogeneity
\cite{lsy2}, would in fact suffice.  The case of a simple box with
hard walls is realized by taking $s=\infty$, while the usual harmonic approximation is $s=2$. It is understood that the
lengths associated with the ground states of $-d^2/dz^2+V(z)$ and
$-(\nabla^\perp)^2+V^\perp(\x^\perp)$ are both of the order $1$ so that
$L$ and $r$ measure, respectively, the longitudinal and the transverse
extensions of the trap.  We denote the ground state energy of
(\ref{3dham}) by $E^{\rm QM}(N,L,r,a)$ and the ground state particle
density by $\rho^{\rm QM}_{N,L,r,a}(\x)$.

While the 3D density is always assumed to be low (in the sense that
distance between particles is large compared to the 3D
scattering length) the 1D density can be either high or
low.  In contrast to 3D gases, {\it
high density} in 1D corresponds to {\it weak interactions}
and vice versa \cite{LL}.

To keep the discussion simple let us
first think of the case that the particles are confined in a box with
dimensions $r$ and $L$.  The 3D particle density is
then $\rho^{\rm 3D}= N/(r^2 L)$ and the 1D density
$\rho^{\rm 1D}=N/L$.  The case of quadratic or more general
homogeneous trapping potentials will be considered later.  
We begin by
describing the division of the space of parameters into two basic
regions.  This decomposition will eventually be refined into five regions, but
for the moment let us concentrate on the basic dichotomy.

In earlier work  \cite{lsy1,lsy2} 
we  proved that the 3D
 Gross-Pitaevskii formula for the energy (including its limiting
 `Thomas-Fermi' case) is correct to leading order in situations in
 which $a$ is small and $N$ is large.  This energy has two parts: The
 energy necessary to confine the particles in the trap, which is
 roughly $N (r^{-2} +L^{-2})$, plus the internal energy of
 interaction, which is $N 4\pi a \rho^{\rm 3D}$.  The
 trouble is that while this formula is correct for a {\it fixed}
 confining potential in the limit $N\to \infty$ with $a^3
 \rho^{\rm 3D}\to 0$, it does not hold uniformly if $r/L$ gets small
 as $N$ gets large.  In other words, new physics can come into play as
 $r/L\to 0$ and it turns out that this depends on the ratio of $a/r^2$
 to $\rho^{\rm 1D}=N/L$ .  There are two basic regimes to 
 consider in highly elongated traps, i.e., when 
 $r \ll L$. They are
 are
 \begin{itemize}
\item The 1D limit of the 
3D Gross-Pitaevskii/`Thomas-Fermi' regime
\item The `true' 1D regime.
\end{itemize}
 The former turns out to be characterized by $aL/r^2N\to 0$, while in
 the latter regime $aL/r^2N$ is of the order one or even tends to
 infinity. (This is usually referred to as the Girardeau-Tonks region,
 but it was Girardeau \cite{gir}, not Tonks, who understood how to
 calculate the states of the 1D hard-core gas.)  These two situations
 correspond to high 1D density (weak interaction) and low 1D density
 (strong interaction), respectively. Physically, the main difference is
 that in the strong interaction regime the motion of the particles in
 the longitudinal direction is highly correlated, while in the weak
 interaction regime it is not. Mathematically, this distinction also
 shows up in our proofs.
 
 In both regions the internal energy of the gas is small compared to
 the energy of confinement which is of the order $N/r^2$.  However,
 this in itself does not imply a specifically 1D behavior.  (If $a$ is
 sufficiently small it is satisfied in a trap of any shape.)  1D
 behavior, when it occurs, manifests itself by the fact that the
 transverse motion of the atoms is uncorrelated while the longitudinal
 motion is correlated (very roughly speaking) in the same way as
 pearls on a necklace.  Thus, the true criterion for 1D behavior is
 that $aL/r^2N$ is of the order unity or larger,  and not merely the condition
 that the energy of confinement dominates the internal energy.

In parallel with the 3D Hamiltonian we consider the 
Hamiltonian for $n$  Bosons in 1D
with delta interaction
and coupling constant $g\geq 0$ , i.e., 
\beq\label{13}
H_{n,g}^{\rm 1D}=\sum_{j=1}^n-\partial^2/\partial z_{j}^2  
+ g \sum_{1\leq i<j\leq n} 
\delta(z_i-z_j)\ .
\eeq
We consider this 
Hamiltonian 
for the $z_{j}$ in
an interval of length $\ell$ in the 
thermodynamic limit, $\ell\to\infty$, $n\to\infty$ with $\rho=n/\ell$ 
fixed. 
The ground state energy per particle in this limit is independent of 
boundary conditions and can, according to \cite{LL}, be written as 
\beq \label{1dendens}
e_{0}^{\rm 1D}(\rho)=\rho^2e(g/\rho) \ ,
\eeq
with a function $e(t)$ determined by a certain 
integral equation. Its asymptotic form is $e(t)\approx 
\half t$ 
for $t\ll 1$ and $e(t)\to \pi^2/3$ for $t\to \infty$. Thus
\beq\label{e0limhigh}
e_{0}^{\rm 1D}(\rho)\approx \half g\rho\ \ \hbox{\rm for}\ \  g/\rho\ll 
1
\eeq
and
\beq\label{e0limlow}
e_{0}^{\rm 1D}(\rho)\approx (\pi^2/3)\rho^2\ \ 
\hbox{\rm for}\ \
g/\rho\gg
1\ .
\eeq

Taking $\rho e_{0}^{\rm 1D}(\rho)$ as a local energy density for an 
inhomogeneous 1D system we can form the energy functional 
\beq\label{genfunc}
\E[\rho]=\int_{-\infty}^{\infty} \!\!\!\! \left( |\nabla\sqrt\rho(z)|^2 +
V_{L}(z)\rho(z) + \rho(z)^3 e(g/\rho(z)) \right) dz
\eeq
with ground state energy defined to be
\beq\label{genfuncen}
E^{\rm 1D}(N,L,g)=\inf \left\{ \E[\rho] \, : \, \rho(z)\geq 0 , \, \int_{-\infty}^{\infty}
\rho(z)dz = N \right\} .
\eeq
By standard methods (cf., e.g., \cite{lsy1}) one can show that there 
is a 
unique minimizer,  i.e., a density 
$\rho_{N,L,g}(z)$ with $\int \rho_{N,L,g}(z)dz=N$ and 
$\E[\rho_{N,L,g}]=
E^{\rm 1D}(N,L,g)$. We define the {\it mean 1D density} of this minimizer to be
\beq
\bar\rho= \frac 1N\int_{-\infty}^{\infty}
\left(\rho_{N,L,g}(z)\right)^2 dz \ .
\eeq
In a rigid box, i.e., for $s=\infty$, $\bar \rho$ is simply $N/L$
(except for boundary corrections), but in more general traps it
depends also on $g$ besides $N$ and $L$.  The order of magnitude of
$\bar\rho$ in the various parameter regions will be described below.

Our main result relates the 3D ground state energy of (\ref{3dham}),
$E^{\rm QM}(N,L,r,a)$, to the 1D density functional energy $E^{\rm
1D}(N,L,g)$ in the large $N$ limit with $g\sim a/r^2$ provided $r/L$
and $a/r$ are sufficiently small. To state this precisely, let
$e^\perp$ and $b(\x^\perp)$, respectively, denote the ground state
energy and the normalized ground state wave function of
$-(\nabla^\perp)^2+V^\perp(\x^\perp)$. The corresponding quantities
for $-(\nabla^\perp)^2+V^\perp_{r}(\x^\perp)$ are $e^\perp/r^2$ and
$b_{r}(\x^\perp)=(1/r)b(\x^\perp/r)$. In the case that the trap is a
cylinder with hard walls $b$ is a Bessel function; for a quadratic
$V^\perp$ it is a Gaussian.

Define $g$ by
\beq\label{defg}
g=\frac {8\pi a}{r^2} \int |b(\x^\perp)|^4 d^2\x^\perp={8\pi a}
\int |b_{r}(\x^\perp)|^4d^2\x^\perp.
\eeq
\begin{thm}\label{T1}
Let $N\to\infty$ and simultaneously $r/L\to 0$ and $a/r\to 0$ in such a way that 
$r^2\bar\rho\cdot\min\{\bar\rho,g\}\to 0$. Then
\beq\label{lim}
\lim    \frac {E^{\rm QM}(N,L,r,a)-Ne^\perp /r^2 }{E^{\rm 1D}(N,L,g)} = 1.
\eeq
Moreover, if we  define 
the 1D quantum mechanical density by averaging
over the transverse variables, i.e., 
\beq
\hat\rho^{\rm QM}_{N,L,r,a}(z)\equiv \int \rho^{\rm QM}_{N,L,r,a}
(\x^\perp,z)d^2\x^\perp \ ,
\eeq
then $\hat\rho^{\rm QM}_{N,L,r,a}(z)/\rho_{N,L,g}(z) \to 1$ in a 
suitable sense.
\end{thm}
Note that because of (\ref{e0limhigh}) and (\ref{e0limlow}) the condition 
$r^2\bar\rho\cdot 
\min\{\bar\rho,g\}\to 0$ is the same as 
\beq \label{condition}e_{0}^{\rm 1D}(\bar \rho)\ll 1/r^2,\eeq
i.e., the average energy per particle associated with the longitudinal
motion should be much smaller than the energy gap between the ground
and first excited state of the confining Hamiltonian in the transverse
directions. Thus, the basic physics is highly quantum-mechanical and has no
classical counterpart. The system can be described by a 
1D functional (\ref{genfunc}), {\it even though the 
transverse trap dimension is much larger than the range of the atomic 
forces.} 

The two regimes mentioned previously correspond to specific
restrictions on the size of the ratio $g/\bar\rho$ as $N\to\infty$,
namely $g/\bar\rho\ll 1$ for the limit of the 3D Gross-Pitaevskii
regime (weak interaction/high density), and $g/\bar\rho>0$ for the
`true' 1D regime (strong interaction/low density). The precise meaning of $\ll$ is that the ratio of the
left side to the right side tends to zero in the limit considered.

We shall now briefly describe the finer division of these two regimes
into five regions altogether. Three of them (Regions 1--3) belong to
the weak interaction regime and two (Regions 4--5) to the strong
interaction regime. In each of these regions the general functional
(\ref{genfunc}) can be replaced by a different, simpler functional,
and the energy $E^{\rm 1D}(N,L,g)$ in the theorem by the ground state
energy of that functional, in analogy with the fact that in 3D the TF
functional, i.e., the GP functional without gradient term, is a limit of
the full GP functional. We state the results
for the ground state energy, but corresponding results hold for the
density. We always assume $N\to \infty$ and $r/L\to 0$.

The five regions are
\medskip

\noindent $\bullet$
{\bf Region 1, the Ideal Gas case:}  $g/\bar\rho\ll N^{-2}$, 
with $\bar\rho\sim N/L$,
corresponding to the trivial case where the interaction
is so weak that it effectively vanishes in the large $N$ limit and
everything collapses to the ground state of $-d^2/dz^2+V(z)$ with 
ground state energy  $e^{\parallel}$.
The energy $E^{\rm 1D}$ in (\ref{lim})
can be replaced by $N  e^{\parallel} /L^2 $. Note that $g/\bar\rho\ll 
N^{-2}$ means that the 3D interaction energy  $\sim \rho^{\rm 3D}a \ll 1/L^2$.
\medskip

\noindent $\bullet$
{\bf Region 2, the 1D GP case:} 
$g/\bar\rho\sim N^{-2}$, with $\bar\rho\sim N/L$, described by a 1D 
Gross-Pitaevskii energy functional 
\beq\label{GPfunct}
\E^{\rm GP}[\rho]=\int_{-\infty}^\infty \left( |\nabla\sqrt\rho(z)|^2+ 
V_L(z)\rho(z) + \half g\rho(z)^2 \right) dz
\eeq
corresponding to the high density approximation (\ref{e0limhigh}) of the 
interaction energy in (\ref{genfunc}). Its ground state energy is
$E^{\rm GP}(N,L,g) = NL^{-2} E^{\rm GP}(1,1,NgL)$, by scaling.
\medskip

\noindent $\bullet$ {\bf Region 3, the 1D TF case:} $N^{-2}\ll
g/\bar\rho \ll 1$, with $\bar\rho\sim (N/L) (NgL)^{-1/(s+1)}$, where
$s$ is the degree of homogeneity of the longitudinal confining
potential $V$.  This region is described by a Thomas-Fermi type
functional \beq\label{TFfunct} \E^{\rm TF}[\rho]=\int_{-\infty}^\infty
\left( V_L(z)\rho(z) + \half g\rho(z)^2 \right) dz \ .  \eeq It is a
limiting case of Region 2 in the sense that $NgL\sim
NaL/r^2\to\infty$, but $a/r$ is sufficiently small so that $g/\bar\rho
\sim (aL/Nr^2)(NaL/r^2)^{1/(s+1)}\to 0$, i.e., the high density
approximation in (\ref{e0limhigh}) is still valid.  In this limit the
gradient term in (\ref{GPfunct}) becomes vanishingly small compared to
the other terms.  The explanation of the factor
$(NaL/r^2)^{1/(s+1)}\sim (NgL)^{1/(s+1)}$ is as follows: The linear
extension $\bar L$ of the minimizing density $\rho^{\rm GP}_{N,L,g}$
is for large values of $NgL$ determined by $V_{L}(\bar L)\sim g(N/\bar
L)$, which gives $\bar L\sim (NgL)^{1/(s+1)} L$.  In addition
condition (\ref{condition}) requires $g\bar\rho \ll r^{-2}$, which
means that $Na/L(NaL/r^2)^{1/(s+1)}\to 0$.  The minimum energy of
(\ref{TFfunct}) has the scaling property $E^{\rm TF}(N,L,g) =
NL^{-2}(NgL)^{s/(s+1)} E^{\rm TF}(1,1,1)$.  \medskip

\noindent $\bullet$
{\bf Region 4, the LL case:}  $g/\bar\rho\sim 1$ , with
$\bar\rho\sim  (N/L )
N^{-2/(s+2)}$, described by an energy functional 
\beq\label{llfunct}
\E^{\rm LL}[\rho]=\int_{-\infty}^\infty \left( V_{L}(z)\rho(z) + \rho(z)^3 
e(g/\rho(z)) \right) dz \ .
\eeq
This region corresponds to the case $g/\bar\rho\sim 1$, so that
neither the high density (\ref{e0limhigh}) nor the low density
approximation (\ref{e0limlow}) is valid and the full LL energy
(\ref{1dendens}) has to be used. The extension $\bar L$ of the system
is now determined by $V_L(\bar L)\sim (N/\bar L)^2$ which leads to
$\bar L_{\rm LL}=L N^{2/(s+2)}$, in contrast to $\bar L_{\rm TF}= L
(NgL)^{1/(s+1)}$ in the TF case. Condition (\ref{condition}) means in
this region that $Nr/\bar L_{\rm LL}\sim N^{s/(s+2)}r/L\to 0$. Since
$Nr/\bar L_{\rm LL}\sim(\bar\rho/g)(a/r)$, this condition is
automatically fulfilled if $g/\bar\rho$ is bounded away from zero and
$a/r\to 0$.  The ground state energy of (\ref{llfunct}), $E^{\rm
LL}(N,L,g)$, is equal to $N\gamma^2 E^{\rm LL}(1,1,g/\gamma)$, where
we introduced the density parameter
\beq
\gamma\equiv  N/\bar L_{\rm LL}=(N/L) N^{-2/(s+2)}\ .
\eeq

\medskip
\noindent $\bullet$
{\bf Region 5, the GT case:} $g/\bar\rho\gg 1$, 
with $\bar\rho\sim  (N/L) N^{-2/(s+2)}$, described by a functional 
with energy density $\sim \rho^3$, corresponding to the
Girardaeu-Tonks limit of the the LL energy density. 
It corresponds to impenetrable particles, i.e, the limiting 
case $g/\bar\rho\to\infty$ and hence formula 
(\ref{e0limlow}) for the energy density. As in Region 4, the mean 
density is here
 $\bar\rho\sim  \gamma$.
The energy functional is
\beq
\E^{\rm GT}[\rho]=\int_{-\infty}^\infty \left( V_{L}(z)\rho(z) + 
({\pi^2}/3 )\rho(z)^3  \right) dz \ ,
\eeq
with minimum energy  $E^{\rm GT}(N,L)= N \gamma^2 E^{\rm GT}(1,1)$.

\medskip
We note that the condition $g/\bar\rho\sim 1$ means that Region 4 
requires the gas cloud to have  aspect ratio $r/\bar L$ of the order 
$N^{-1}(a/r)$ or smaller, where $\bar 
L\sim LN^{2/(s+2)}$ is the length of the cloud. Experimentally, such 
small aspect ratios are quite a 
challenge and the situations described in \cite{bongs, goerlitz, 
greiner, schreck} 
are still  rather far from this regime. It may not be completely out of 
reach, however.

Regions 1--3 can be reached as limiting cases of a 3D
Gross-Pitaevskii theory.  In this sense, the behavior in these regions
contains remnants of the 3D theory, which also shows up in the the
fact that Bose-Einstein condensation (BEC) prevails in Regions 1 and 2 
\cite{LSY}. 
Heuristically, these traces
of 3D can be understood from the fact that in Regions 1--3 the 1D
formula for energy per particle, $g\rho\sim aN/(r^2 L)$, gives the same
result as the 3D formula \cite{LY1998}, i.e., scattering length times
3D density.  This is no longer so in Regions 4 and 5 and different
methods are required.

BEC probably occurs in part of Region
3, besides Regions 1 and 2, but we
cannot prove this and it remains an open problem. BEC means
\cite{LS} that the one-body density matrix $\gamma(\x, \x')$
factorizes as $N\phi(\x)\phi(\x')$ for some normalized $\phi$. This,
in fact, is 100\% condensation and this is what we prove occurs (in
the $N\to \infty$ limit, of course). The function $\phi $ is the
square-root of the minimizer of the 1D GP functional (\ref{GPfunct})
times the transverse function $b_r(\x^\perp)$. The proof is similar to
the work in
\cite{LS}. 
BEC is not expected in Regions 4 and 5. Lenard \cite{Lenard} showed
that the largest eigenvalue of $\gamma$ grows only as $N^{1/2}$ for a
homogeneous gas of 1D impenetrable bosons and, according to
\cite{papenbrock} and \cite{forrester}, this holds also for a GT gas
in a harmonic trap. (The exponent 0.59 in \cite{girardeau} can
probably be ascribed to the small number of particles ($N=10$)
considered.)

We now outline the main steps in the proof of the 
Theorem, referring to \cite{LSY} for full details.
The different parameter regions have to be treated by different 
methods, a watershed lying between Regions 1--3 on the one hand 
and Regions 4--5 on the other. In Regions 1--3, 
similar methods as in the proof of the 3D Gross-Pitaevskii limit 
theorem in \cite{lsy1} can be used. This 3D proof needs  considerable 
modifications, however,  because in \cite{lsy1} the external 
potential is fixed and the estimates are not uniform in the ratio $r/L$.

To prove (\ref{lim}) one has to establish upper and lower bounds, with
controlled errors, on the QM many-body energy in terms of the energies
obtained by minimizing the energy functionals appropriate for the
various regions.  The limit theorem for the densities can be derived
from the energy estimates in a standard way by variation with respect
to the external potential $V_{L}$.  As usual, the upper bounds for the
energy are easier than the lower bounds, but nevertheless not simple,
in particular not for `hard' potentials $v$.

The upper bound in Regions 1--3 is obtained from a variational ansatz
of the form
$
\Psi(\x_1,\dots,\x_N)=F(\x_1,\dots,\x_N) \hbox{$\prod_{k=1}^N$}
b_r(\x_k^\perp)\sqrt{\rho^{\rm GP}}(z_{k}),
$
with
$
F(\x_1,\dots,\x_N)$ $=\prod_{k=1}^N f(\x_{k}-\x_{j(k)})
$
where $\x_{j(k)}$ is the nearest neighbor of $\x_{k}$ among the points 
$\x_{j}$, $j<k$,
$b_r(\x^\perp)$  is the lowest eigenfunction of $-(\nabla^\perp)^2+V_{r}^\perp$,
and ${\rho^{\rm GP}(z)}$ the minimizer of the 1D GP functional 
(\ref{GPfunct}). The function $f$ is, up to a cut-off length that 
has to be chosen optimally, the zero energy 
scattering solution for the two-body Hamiltonian with interaction $v_{a}$. 
The form of $F$, inspired by \cite{dyson},  is chosen rather than a Jastrow ansatz 
$\prod_{i<j}f(\x_{i}-\x_{j})$ because it is computationally 
simpler for the purpose of obtaining rigorous estimates. 

For an upper bound in Regions 4--5 a natural variational ansatz would 
 appear to be
$\Psi(\x_1,\dots,\x_N)=F(\x_1,\dots,\x_N) \hbox{$\prod_{k=1}^N$} b_r(\x_k^\perp)
\psi(z_{1},\dots,z_{N})$ 
where $\psi$ is the ground state of $H^{\rm 1D}_{N,g}$ with the
external potential $V_{L}$ added.  However, in order to make a link
with the exact solution (\ref{1dendens}) for a homogeneous gas, but
also to control the norm of the trial function, it turns out to be
necessary to localize the particles by dividing the trap into finite
`boxes' (finite in $z$-direction), with a finite particle number in
each box and making the ansatz  with
the boundary condition $\Psi=0$ for each box individually.  The particles
are then distributed optimally among the boxes to minimize the energy. 
This box method, but with the boundary condition $\nabla\Psi=0$, is also used
for the lower bounds to the energy.  Another essential device for the lower
bounds is {\it Dyson's Lemma} that was also used in \cite{LS, lsy1,LY1998}.  This lemma, which goes back to Dyson's
seminal paper \cite{dyson} on the hard core Bose gas, estimates the
kinetic and potential energy for a Hamiltonian with a `hard' potential $v$ of finite range from below
by the potential energy of a `soft' potential $U$ of larger range but 
essentially the same scattering length as $v$. 
Borrowing a tiny part of the kinetic energy it is then possible to do
perturbation theory with the `soft' potential $U$  and use Temple's inequality
\cite{TE} to bound the errors.  A direct application of perturbation
theory to the original potential $v$, on the other hand, is in general
not possible.

A core lemma for Regions 4--5 is a lower bound on the 3D ground state energy in a finite box in 
terms of the 1D energy of the Hamiltonian (\ref {13}), both with 
the boundary condition $\nabla\Psi=0$. Denoting the former energy by 
$E_{{\rm box}}^{\rm 3D}$ and the 
latter by $E_{{\rm box}}^{\rm 1D}$, this bound for $n$ particles in 
a box of length $\ell$ in the $z$-direction reads
\begin{equation*}
E_{{\rm box}}^{\rm 3D}-\frac{ne^\perp}{r^2} \geq E_{{\rm box}}^{\rm
1D}
\left( 1 -C n 
\left(\frac{a}{r}\right)^{1/8}\left[1+\frac {nr}{\ell}
\left(\frac{a}{r}\right)^{1/8}   \right]\right)  \ 
\end{equation*}
with a constant $C$. To prove this bound the ground state 
wave function 
is first written as a product  of  $\prod_{k} b_r(\x_k^\perp)$ and a 
function $G(\x_{1},\dots,\x_{n})$. This subtracts 
${ne^\perp}/{r^2}$ from $E_{{\rm box}}^{\rm 3D}$ but the resulting minimization 
problem for $G$ involves the weighted measure $\prod_{k} 
b_r(\x_k^\perp)^2 d^3\x_{k}$ in place of $\prod_{k} 
d^3\x_{k}$. Nevertheless, Dyson's Lemma can be 
used, and after the `hard' potential $v$ has been replaced by the `soft' 
potential $U$, it is possible to integrate the transverse variables away 
and obtain a minimization problem for a 1D many body Hamiltonian with 
interaction $d(z_{i}-z_{j})=a\int\int b_r(\x_{i}^\perp)^2b_r(\x_{j}^\perp)^2 
U(\x_{i}-\x_{j})d^2\x_{i}^\perp d^2\x_{j}^\perp$. In the limit considered 
this 
converges to a delta interaction with the coupling constant 
(\ref{defg}). The error terms in the estimate for $E_{\rm box}^{\rm 3D}$ arise 
both from Temple's inequality and the replacement of $d$ by a delta 
function, among other things. When the particles are distributed optimally among the boxes 
to obtain a global lower bound, superadditivity of the energy and 
convexity of the energy density $\rho^3 e(g/\rho)$ are used, 
generalizing corresponding arguments in \cite{LY1998}.

In conclusion, we have reported a rigorous analysis of the parameter
regions in which a Bose gas in an elongated trap may or may not be
expected to display 1D behavior in its ground state. This takes the
form of theorems about the ground state energy and density. We always
consider $N$ to be large and $r/L$ to be small, but this is not enough
to distinguish 1D from 3D behavior. The distinction occurs when,
additionally, the 1D density is not too large, e.g., $aL/r^2 \sim
N^{3/2}$ in a harmonic trap. The 1D behavior, when it occurs, is
described by the delta-function gas solved in \cite{LL}. We also
present a 1D energy functional, analogous to the Gross-Pitaevskii
functional, that correctly describes the energy and density in all the
5 parameter regions considered here.

\end{document}